\documentclass[doublecol]{epl2} 

\usepackage{amsmath,amssymb}

\title{Nonlinear dynamics analysis of a low-temperature-differential kinematic Stirling heat engine}
\shorttitle{Nonlinear dynamics analysis of a low-temperature-differential kinematic Stirling heat engine} 

\author{Yuki Izumida}
\shortauthor{Yuki Izumida}

\institute{                    
 Department of Complex Systems Science, Graduate School of Informatics, Nagoya University, Nagoya 464-8601, Japan
}

\abstract{
The low-temperature-differential (LTD) Stirling heat engine technology constitutes one of the important sustainable energy technologies. 
The basic question of how the rotational motion of the LTD Stirling heat engine is maintained or lost based on the temperature difference is thus
a practically and physically important problem that needs to be clearly understood.
Here, we approach this problem by 
proposing and investigating a minimal nonlinear dynamic model of an LTD kinematic Stirling heat engine.
Our model is described as a driven nonlinear pendulum where the motive force is the temperature difference.
The rotational state and the stationary state of the engine are described as a stable limit cycle and a stable fixed point of the dynamical equations, respectively.
These two states coexist under a sufficient temperature difference, whereas the stable limit cycle does not exist under a temperature difference that is too small.
Using a nonlinear bifurcation analysis, we show that 
the disappearance of the stable limit cycle occurs 
via a homoclinic bifurcation, with the temperature difference being the bifurcation parameter.}

\begin{document}

\maketitle

\section{Introduction}  
Technology that can use thermal energy 
as a resource is one of the most important technologies needed to develop a sustainable society.
The Stirling heat engine, which was invented in the early 19th century, 
has received renewed attention over the past several years for its potential application in sustainable energy technology~\cite{SS,KW}.
Owing to technological innovations 
during the past few decades, some types of the Stirling heat engine that can operate under very low temperature differences have been developed~\cite{S}, 
and make it possible for the use of thermal energy in our daily life.
The demonstration of a low-temperature-differential (LTD) Stirling heat engine operating in the palm of the hand 
under the difference between the body temperature and room temperature is remarkable~\cite{S}, and serves as useful material for both teaching and learning thermodynamics. 
\begin{figure}
\begin{center}
\includegraphics[scale=0.34]{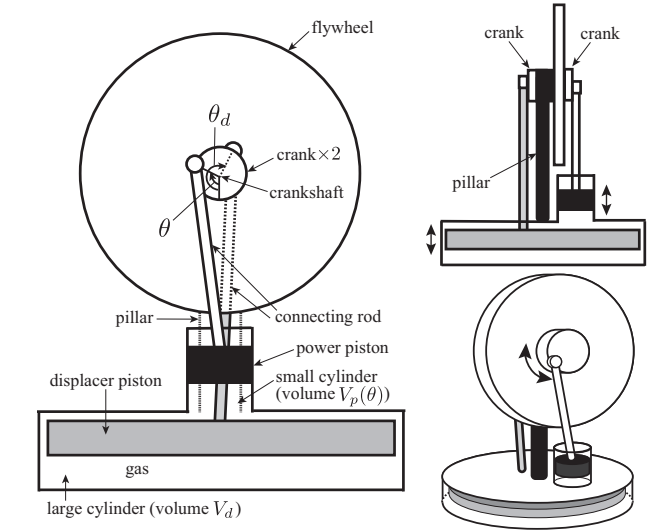}
\caption{Schematic illustration of the low-temperature-differential (LTD) kinematic Stirling heat engine: Front view (left), side view (right top), and oblique view (right down).}\label{stirling_engine}
\end{center}
\end{figure}

From a fundamental physics perspective, it is important to understand how the rotational motion of the LTD Stirling heat engine 
is maintained or lost based on the temperature difference.
In contrast to the ideal Stirling thermodynamic cycle with an external operation~\cite{SS,KW}, 
actual Stirling heat engines run autonomously in a self-sustained manner under a sufficient temperature difference. 
This is accomplished with the aid of a crank-piston mechanism~\cite{SS,KW}.
Mathematical modeling of the Stirling heat engine thus involves solving coupled equations that describe both
the dynamics of a working fluid and that of the mechanical degrees of freedom of the engine~\cite{CB}.
While a simple theoretical model of an LTD Stirling heat engine that reproduces the rotational motion of the engine 
has been proposed~\cite{RGK}, how this motion is lost with decreasing temperature difference remains to be clarified.

In this study, we provide an answer to this question by proposing and investigating a minimal model of an LTD Stirling heat engine based on nonlinear dynamics.
To be more specific, we model a kinematic Stirling heat engine that has been adopted to implement actual LTD Stirling heat engines~\cite{KW}.
Our model is the simplest in the sense that dynamical equations with the fewest degrees of freedom 
and a stable limit cycle solution interpreted as the rotational motion of the engine 
are involved.
We then analyze a bifurcation of the stable limit cycle solution, with the temperature difference being the bifurcation parameter.

\begin{figure}
\begin{center}
\includegraphics[scale=0.35]{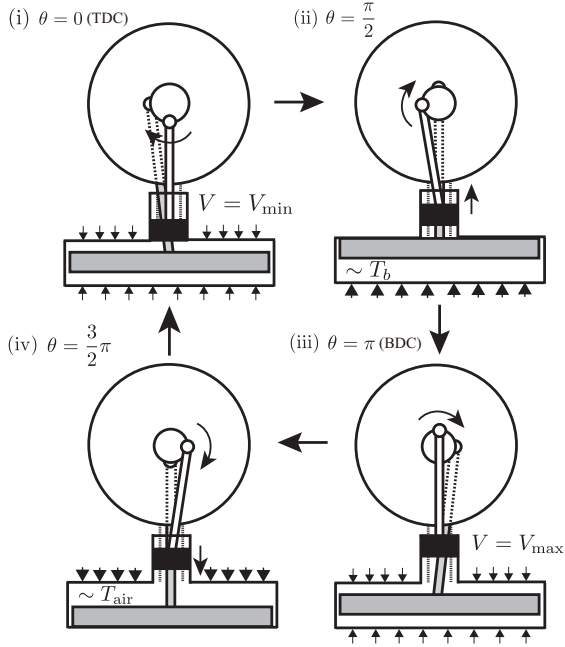}
\caption{Schematic illustration of the cycle process of the LTD kinematic Stirling heat engine for $\Delta {T}>0$ (clockwise rotation):
(i) $\to$ (ii) transfer stroke, (ii) $\to$ (iii) expansion stroke, (iii) $\to$ (iv) transfer stroke, and (iv) $\to$ (i) compression stroke. 
}\label{stirling_explanation}
\end{center}
\end{figure}

\section{Kinematic Stirling heat engine} 
The kinematic Stirling heat engine is 
a gamma-configuration (cylinder-split) Stirling heat engine~\cite{SS,KW}.
In fig.~\ref{stirling_engine}, we illustrate
a kinematic Stirling heat engine designed for an LTD operation (a prototype of this LTD kinematic engine is the N-92 type designed and built by Senft~\cite{S}): 
The working substance, a gas, is confined into a cylinder with total volume $V$.
The cylinder consists of a large cylinder with volume $V_d$ and a small cylinder with volume $V_p$, where $V=V_d+V_p$.
The small cylinder with a power piston on its top is placed on the large cylinder. 
The power piston is connected to a crank with a small radius via a connecting rod.
The reciprocating motion of the power piston is converted to a rotational motion by the crank.
The cylinder is separated into upper and lower regions by a reciprocating displacer piston in the large cylinder. 
The displacer piston is connected to the other crank with the same radius as the former 
via a connecting rod. The two cranks, along with a flywheel with a large moment of inertia, are combined to rotate as a single synchronized unit with the same crankshaft centered at the cranks.
The phase angle $\theta_d$ (${\rm mod} \ 2\pi$) of the crank connected to the displacer piston is arranged 
such that it leads the phase angle $\theta$ (${\rm mod} \ 2\pi$) of the crank connected to 
the power piston by $\frac{\pi}{2}$ as $\theta_d=\theta+\frac{\pi}{2}$~\cite{KW}, where $\theta_d$ and $\theta$ increase clockwise from the origin corresponding to the lowest point of each piston. 
This unit, which is supported by a pillar, is fixed above the cylinder.
There is a small space of negligible volume between the displacer piston and the wall of the large cylinder. 
The gas is transferred back and forth through the space between the upper and the lower regions of the cylinder because of the reciprocating motion of the displacer piston.
The volume $V_d$ of the large cylinder that agrees with the swept volume of the displacer piston remains constant, while
that of the small cylinder $V_p$ changes with $\theta$.
At the bottom of the large cylinder, the engine contacts a heat reservoir (e.g., a warm palm or cold ice) at temperature $T_b$. 
The surrounding air with temperature $T_{\rm air}$ serves as the other heat reservoir
and the atmospheric pressure $p_{\rm air}$ acts on the power piston.
As shown in fig.~\ref{stirling_engine}, an LTD Stirling heat engine has some characteristics in its geometry 
that include the following~\cite{KW,S}.
The compression ratio of the engine is small;
the diameter of the large cylinder and the displacer piston is large;
the displacer piston is short, and its stroke is small;
the bottom and top surfaces of the large cylinder for thermal conduction is large.

For $\Delta T\equiv T_b-T_{\rm air}>0$, the kinematic Stirling heat engine runs clockwise.
Initially, the phase angle of the crank is located at $\theta=0$ ($\theta_d=\frac{\pi}{2}$) with a positive angular velocity.
The points $\theta=0$ and $\theta=\pi$ where the volume of the cylinder reaches its minimum and maximum 
are called the top dead center (TDC) and the bottom dead center (BDC), respectively~\cite{KW,MCHGAS};  
a torque force transmitted from the power piston to the crank vanishes at these points.
The engine then undergoes the following cyclic process (fig.~\ref{stirling_explanation})~\cite{SS,KW}. (i) $\to$ (ii) transfer stroke: 
The crank rotates clockwise and overcomes $\theta=0$ (TDC) with the aid of the moment of inertia of the flywheel even against a frictional torque,
and the gas expands from the minimum volume $V=V_{\rm min}=V_d$ at $\theta=0$ to push the power piston against the atmospheric pressure. 
Meanwhile, the displacer piston transfers most of the gas to the lower region of the cylinder.
The more gas that gathers in the lower region, the more it is heated up by the hot heat reservoir (warm palm) with $T_b$ at the bottom surface of the large cylinder.
(ii) $\to$ (iii) expansion stroke: 
The gas temperature 
becomes close to $T_b$ at $\theta=\frac{\pi}{2}$ ($\theta_d=\pi$) 
where the gas is completely separated from the cold heat reservoir (surrounding air) at the top surface of the large cylinder.
The gas continues to expand and pushes the power piston until the volume reaches the maximum $V=V_{\rm max}$ at $\theta=\pi$ ($\theta_d=\frac{3}{2}\pi$).
(iii) $\to$ (iv) transfer stroke: 
With the aid of the moment of inertia of the flywheel, the crank keeps rotating, overcoming $\theta=\pi$ (BDC), and the power piston 
starts to compress the gas against the gas pressure.
Meanwhile, the displacer piston transfers the gas to the upper region of the large cylinder.
The more gas that gathers in the upper region, the more it is cooled down by the cold heat reservoir with $T_{\rm air}$ at the top surface of the large cylinder.
(iv) $\to$ (i) compression stroke:
The gas temperature becomes close to $T_{\rm air}$ at $\theta=\frac{3}{2}\pi$
($\theta_d=2\pi$) where the gas is completely separated from the hot heat reservoir at the bottom surface of the large cylinder.
The power piston continues to compress the gas until the volume reaches the minimum $V=V_{\rm min}=V_d$ at $\theta=2\pi$ ($\theta_d=\frac{\pi}{2}$)
and the engine returns to the initial state.
In this cyclic process, 
the displacer piston serves as an automatic controller that realizes the alternate attachment and detachment of the heat reservoirs with the gas, depending on its location.
In contrast, the power piston is a motive part that converts the thermal energy into the rotational motion of the engine via the crank.

This engine runs counterclockwise for $\Delta T <0$~\cite{S}. 
The engine starts from $V=V_{\rm min}=V_d$ at $\theta=0$ with a negative angular velocity and follows the following sequence: (i) $\to$ (ii) transfer stroke, (ii) $\to$ (iii) expansion stroke, (iii) $\to$ (iv) transfer stroke, and (iv) $\to$ (i) compression stroke, and this sequence is the same as the clockwise case with $\Delta T>0$ in fig.~\ref{stirling_explanation}. 
One difference is that the gas is transferred to the upper region of the cylinder at the first transfer stroke and 
is heated up by the surrounding air (which acts as the hot heat reservoir) while it is transferred to the lower region of the cylinder at the second transfer stroke and is cooled down by, e.g., cold ice acting as the cold heat reservoir.

\section{Nonlinear dynamic model}
We propose a minimal nonlinear dynamic model that captures the above features of the LTD kinematic Stirling heat engine.
We assume that the gas, the working substance, can be approximated as an ideal gas.
We also assume that the gas has spatially uniform temperature $T$ and density
over the entire cylinder with total volume $V$. 

Our model is a coupled system with mechanical and thermodynamic degrees of freedom~\cite{CB,RGK}.
The mechanical degree of freedom is the phase angle $\theta$ of the crank, which has a radius $r$.
We assume that the equation of motion of the crank is given by 
\begin{eqnarray}
I \frac{d^2 \theta}{dt^2}=r F \sin \theta-\Gamma \frac{d\theta}{dt},\label{eq.wheel}
\end{eqnarray}
where $F\equiv \sigma_p\left(\frac{nRT}{V(\theta)}-p_{\rm air}\right)$.
Here, $I$ is the moment of inertia of the crank with the flywheel attached, 
$V_p(\theta)\equiv r(1-\cos \theta)\sigma_p$ and $V_d=2r \sigma_d$ are the volume of the small cylinder and that of the large cylinder, respectively, 
with $\sigma_p$ and $\sigma_d$ being the bottom areas of these cylinders. 
The total volume $V(\theta)$ of the cylinder is given by 
\begin{eqnarray}
V(\theta)=V_d+V_p(\theta)=2r \sigma_d+r(1-\cos \theta)\sigma_p,\label{eq.total_volume}
\end{eqnarray}
where $V_{\rm min}=V_d=2r\sigma_d$ at $\theta=0$ and $V_{\rm max}=V_{\rm min}+2r\sigma_p$ at $\theta=\pi$.
$F$ is a net force acting on the power piston created by the pressure difference between the internal gas and the surrounding air.
The gas pressure is given by the equation of state for the ideal gas $\frac{nRT}{V(\theta)}$ with $n$ and $R$ being the number of moles of the gas and the molar gas constant, respectively.
$\Gamma$ is a coefficient of the frictional torque, and the mass of the displacer piston, mass of the connecting rods, and gravity are negligible for simplicity.
We can derive eq.~(\ref{eq.wheel}) from the standard setup of a crank-piston mechanism~\cite{MCHGAS} by imposing conditions $\frac{r}{l}\ll 1$ and $\frac{m_p r^2}{I}\ll 1$,
where $l$ and $m_p$ are the length of the connecting rod and the mass of the power piston with negligible friction, respectively.
Under the condition $\frac{r}{l}\ll 1$, the net force $F$ acting on the power piston is transmitted to the crank via the connecting rod in a nearly parallel fashion~\cite{CB}.
This allows us to interpret the part $F \sin \theta$ on the r.h.s. of eq.~(\ref{eq.wheel}) as the tangential component of the force applied to the crank
and hence $rF \sin \theta$ as the rotational torque, which vanishes at $\theta=0$ (TDC) and $\theta=\pi$ (BDC).

The thermodynamic degree of freedom in our model is the gas temperature $T$.
The time-evolution equation of $T$ is given by the first law of thermodynamics (the law of energy conservation) for the ideal gas~\cite{R}:
\begin{eqnarray}
\frac{f}{2}nR\frac{dT}{dt}=\sum_{j=b,{\rm air}}G_j(\theta)(T_j-T)-\frac{nRT}{V(\theta)}\frac{dV(\theta)}{dt},\label{eq.1st_law}
\end{eqnarray}
where $f\ge 3$ is the total number of spatial and internal degrees of freedom per molecule and $\frac{dV(\theta)}{dt}=\frac{dV_p(\theta)}{dt}=\sigma_p r \sin \theta \frac{d\theta}{dt}$.
We assume that the heat flows from the two heat reservoirs according to the Fourier law 
and flows only through the top and the bottom surface of the large cylinder into the gas.  
$G_j(\theta)$ ($j=b, {\rm air}$) in eq.~(\ref{eq.1st_law}) is a state-dependent thermal conductance defined by 
\begin{eqnarray}
G_j(\theta)\equiv G\chi_j(\theta).\label{eq.thermal_cond_1}
\end{eqnarray}
Here $G$ is a usual thermal conductance and 
\begin{eqnarray}
\chi_j(\theta)\equiv \frac{V_j(\theta)}{V_d} \ \ (0\le \chi_j(\theta)\le 1) \label{eq.degree_coupling}
\end{eqnarray}
is a function expressing a degree of coupling between the gas 
and the heat reservoir on each side of the large cylinder separated by the displacer piston, where
\begin{eqnarray}
V_b(\theta)\equiv r (1+\sin \theta)\sigma_d, \ 
V_{\rm air}(\theta)\equiv V_d-V_b(\theta) \label{eq.vb_def}
\end{eqnarray}
are the separated volumes of the large cylinder by the displacer piston.
According to eq.~(\ref{eq.degree_coupling}), 
the amount of heat flowing into one side of the large cylinder increases owing to the increase in volume on this side, 
while the amount of heat flowing into the other side of the large cylinder reduces with the decrease in volume on that side.
The degree-of-coupling function eq.~(\ref{eq.degree_coupling}) models the role of the displacer piston as shown 
in the transfer strokes in (i) $\to$ (ii) and (iii) $\to$ (iv) in fig.~\ref{stirling_explanation},
where the reciprocating motion of the displacer piston alternately transfers the gas into one side of the large cylinder, 
making the gas thermally isolated from the heat reservoir on the other side of the large cylinder 
($\chi_b(\frac{\pi}{2})=1$ and $\chi_{\rm air}(\frac{\pi}{2})=0$ correspond to (ii) in fig.~\ref{stirling_explanation} 
and $\chi_b(\frac{3}{2}\pi)=0$ and $\chi_{\rm air}(\frac{3}{2}\pi)=1$ correspond to (iv) in fig.~\ref{stirling_explanation}).

\begin{figure}
\begin{center}
\includegraphics[scale=0.3]{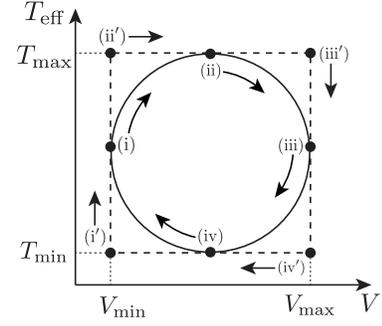}
\caption{$T_{\rm eff}$--$V$ diagram of the kinematic Stirling heat engine in eq.~(\ref{T_eff_2}) (solid curve), 
where $T_{\rm min}=T_{\rm air}$ ($T_b$) for $\Delta T>0$ ($\Delta T<0$) and 
$T_{\rm max}=T_b$ ($T_{\rm air}$) for $\Delta T>0$ ($\Delta T<0$).  
The dashed line indicates the $T$--$V$ diagram of the ideal Stirling thermodynamic cycle consisting of (i$'$) $\to$ (ii$'$) constant-volume heating process ($T=T_{\rm min}\to T_{\rm max}$ at $V=V_{\rm min}$), 
(ii$'$) $\to$ (iii$'$) isothermal expansion process ($V=V_{\rm min}\to V_{\rm max}$ at $T=T_{\rm max}$), 
(iii$'$) $\to$ (iv$'$) constant-volume cooling process ($T=T_{\rm max}\to T_{\rm min}$ 
at $V=V_{\rm max}$), and (iv$'$) $\to$ (i$'$) isothermal compression process ($V=V_{\rm max}\to V_{\rm min}$ at $T=T_{\rm min}$).
}\label{T_protocol}
\end{center}
\end{figure}

To clarify the role of the displacer piston from another perspective, 
it is convenient to rewrite the heat flow term in eq.~(\ref{eq.1st_law}) as
\begin{eqnarray}
\sum_{j=b,{\rm air}}G_j(\theta)(T_j-T)=G(T_{\rm eff}(\theta)-T),\label{eq.one_reservoir}
\end{eqnarray}
where we used eqs.~(\ref{eq.thermal_cond_1})--(\ref{eq.vb_def}), and 
\begin{eqnarray}
T_{\rm eff}(\theta)\equiv T_{\rm air}+\frac{1+\sin \theta}{2}\Delta T\label{T_eff}
\end{eqnarray}
is an effective temperature.
While this engine runs autonomously, we may interpret eq.~(\ref{eq.one_reservoir}) as if one heat reservoir with
$T_{\rm eff}(\theta)$ is tactically attached to the engine by an external agent as in the case of a usual thermodynamic cycle.
We can also rewrite eq.~(\ref{T_eff}) as a function of $V$ by using eq.~(\ref{eq.total_volume}) as
\begin{eqnarray}
T_{\rm eff}(V)\equiv T_{\rm air}+\frac{1\pm \sqrt{1-\left[1-\frac{2(V-V_{\rm min})}{V_{\rm max}-V_{\rm min}}\right]^2}}{2}\Delta T,\label{T_eff_2}
\end{eqnarray}
where the plus (minus) sign applies for $0\le \sin \theta \le 1$ ($-1 \le \sin \theta <0$)
(the solid curve in fig.~\ref{T_protocol}). 
This is comparable to the $T$--$V$ diagram of the ideal Stirling thermodynamic cycle~\cite{SS,KW} (the dashed line in fig.~\ref{T_protocol}).
An important difference is that the ideal cycle has a square shape with a clear distinction between the temperature-changing process and the volume-changing process
whose temperature modulation
is a periodic square-shaped function of $\theta$. 
However, the $T_{\rm eff}$--$V$ diagram of the present kinematic engine has a circular shape, as in eq.~(\ref{T_eff_2}),
whose temperature modulation is the sinusoidal-shaped function of $\theta$ as in eq.~(\ref{T_eff}), which loosely approximates the ideal cycle and is better suited for a practical engine.
We note that essentially the same idea of a state-dependent thermal contact of this kind 
was adopted in earlier studies on different autonomous heat engines~\cite{AGJ,A,RNASS}.

For convenience, we nondimensionalize our coupled equations~(\ref{eq.wheel}) and (\ref{eq.1st_law}).
We define the following nondimensionalized quantities as
$\tilde{t}\equiv \sqrt{\frac{nRT_{\rm air}}{I}}t$,
$\tilde{\omega}\equiv \frac{\omega}{\sqrt{\frac{nRT_{\rm air}}{I}}}$,
$\tilde{G}\equiv \frac{G}{nR \sqrt{\frac{nRT_{\rm air}}{I}}}$, 
$\tilde{\sigma}\equiv \frac{\sigma_p}{\sigma_d}$,
$\tilde{\Gamma}\equiv \frac{\Gamma}{\sqrt{nRT_{\rm air} I}}$,
$\tilde{p}_{\rm air}\equiv \frac{\sigma_d r p_{\rm air}}{nRT_{\rm air}}$, 
$\tilde{T}\equiv \frac{T}{T_{\rm air}}$,
and $\Delta \tilde{T}\equiv \frac{\Delta T}{T_{\rm air}}=\frac{T_b-T_{\rm air}}{T_{\rm air}}$
is a nondimensionalized temperature difference between the heat reservoirs that will serve as a bifurcation parameter.
By defining $\tilde{\omega} \equiv \frac{d\theta}{d\tilde{t}}$ as a nondimensionalized quantity 
of the angular velocity $\omega\equiv \frac{d\theta}{dt}$, 
we can obtain the nondimensionalized dynamical equations corresponding to eqs.~(\ref{eq.wheel}) and (\ref{eq.1st_law}) as
\begin{eqnarray}
&&\frac{d\theta}{d\tilde{t}}=\tilde{\omega},\label{eq.theta_nodim}\\
&&\frac{d\tilde{\omega}}{d\tilde{t}}=\tilde{\sigma}\left(\frac{\tilde{T}}{\tilde{V}(\theta)}-\tilde{p}_{\rm air}\right)\sin \theta-\tilde{\Gamma}\tilde{\omega},\label{eq.ptheta_nodim}\\
&&\frac{d\tilde{T}}{d\tilde{t}}=\frac{2}{f}\tilde{G}\left[\tilde{T}_{\rm eff}(\theta)-\tilde{T}\right]-\frac{2\tilde{\sigma}\tilde{T}\sin \theta}{f\tilde{V}(\theta)}\tilde{\omega},\label{eq._T_nodim}\label{eq.T_nodim}
\end{eqnarray}
where $\tilde{T}_{\rm eff}(\theta)\equiv 1+\frac{1+\sin \theta}{2}\Delta \tilde{T}$ and $\tilde{V}(\theta)\equiv 2+\tilde{\sigma}(1-\cos \theta)$.

For simplicity, 
we make the assumption of a time-scale separation between $\theta$ and $\tilde{\omega}$ as slow variables and $\tilde{T}$ as a fast variable.
Under this assumption,
$\tilde{T}$ in eq.~(\ref{eq.T_nodim}) starting from an initial value at an initial time quickly relaxes to 
an adiabatic solution $\tilde{T}(\theta, \tilde{\omega})$ determined by the slow variables,
which is obtained by solving $\frac{d\tilde{T}}{d\tilde{t}}=0$ in eq.~(\ref{eq.T_nodim})\revision{:}
\begin{eqnarray}
\tilde{T}(\theta,\tilde{\omega})=\frac{\tilde{G}\tilde{T}_{\rm eff}(\theta)}{\tilde{G}+\frac{\tilde{\sigma}\sin \theta \tilde{\omega}}{\tilde{V}(\theta)}}.\label{eq.adiabatic_elimination_T}
\end{eqnarray}
By substituting eq.~(\ref{eq.adiabatic_elimination_T}) into eq.~(\ref{eq.ptheta_nodim}), 
we can eliminate $\tilde{T}$ from eq.~(\ref{eq.ptheta_nodim}) (adiabatic elimination~\cite{H}) and obtain the simple dynamical equations 
closed by the slow variables instead of eqs.~(\ref{eq.theta_nodim})--(\ref{eq.T_nodim}) as follows:
\begin{eqnarray}
&&\frac{d\theta}{d\tilde{t}}=\tilde{\omega},\label{eq.theta_nodim_2}\\
&&\frac{d\tilde{\omega}}{d\tilde{t}}=\tilde{\sigma}\left(\tilde{p}(\theta,\tilde{\omega})-\tilde{p}_{\rm air}\right)\sin \theta-\tilde{\Gamma}\tilde{\omega},\label{eq.ptheta_nodim_2}
\end{eqnarray}
where
\begin{eqnarray}
\tilde{p}(\theta,\tilde{\omega}) \equiv \frac{\tilde{T}(\theta,\tilde{\omega})}{\tilde{V}(\theta)}=\frac{\tilde{G}\tilde{T}_{\rm eff}(\theta)}{\tilde{G}\tilde{V}(\theta)+\tilde{\sigma}\sin \theta \tilde{\omega}}.
\end{eqnarray}
With eqs.~(\ref{eq.theta_nodim_2}) and (\ref{eq.ptheta_nodim_2}), we can consider the dynamics of our engine
on the phase plane $(\theta, \tilde{\omega})$ as a dynamical system.
This model with eqs.~(\ref{eq.theta_nodim_2}) and (\ref{eq.ptheta_nodim_2}) may be regarded as a certain type of a driven nonlinear 
pendulum~\cite{Stz} that is motive-powered by a temperature difference.

\section{Numerical calculations}
We perform numerical calculations of eqs.~(\ref{eq.theta_nodim})--(\ref{eq.T_nodim}) and eqs.~(\ref{eq.theta_nodim_2}) and (\ref{eq.ptheta_nodim_2}).
We choose the parameters in the equations by assuming actual palm-sized LTD Stirling heat engines, some of which have been determined by referring to~\cite{S,RGK}: 
We set $f=3$, $r=0.005 \ {\rm m}$, $I=0.00025 \ {\rm kg \cdot m^2}$, $\sigma_d=0.005 \ {\rm m}^2$, $\sigma_p=0.02\sigma_d=0.0001 \ {\rm m}^2$, $p_{\rm air}=1020 \ {\rm hPa}$, 
$G=45 \ {\rm J\cdot s^{-1} \cdot K^{-1}}$,
$T_b=257 \sim 329 \ {\rm K}$, $T_{\rm air}=293 \ {\rm K}$, $n=0.002031 \ {\rm mol}$, 
and $R=8.314 \ {\rm J \cdot K^{-1} \cdot {mol}^{-1}}$.
These parameters yield the nondimensionalized parameters in the equations as $\tilde{\sigma}=0.02$, $\tilde{G}\simeq 18.94$, $\tilde{p}_{\rm air}\simeq 0.5154$, and
$\Delta \tilde{T}\simeq -0.1229 \sim 0.1229$.
Because an estimation of $\Gamma$ is difficult, here we simply choose 
$\tilde{\Gamma}=0.0025$ as an adjusted value that makes the rotational speed of the engine to be approximately dozens to hundreds of revolutions per minute (rpm) on average,
as observed in actual LTD Stirling heat engines~\cite{S}.
We use the $4$th Runge--Kutta  method with time interval $\Delta \tilde{t}=0.0001$ in the numerical calculations.
Hereafter, we denote by a quantity with $\left<\cdot \right>$ a long-time averaged one.

\begin{figure}
\begin{center}
\includegraphics[scale=0.6]{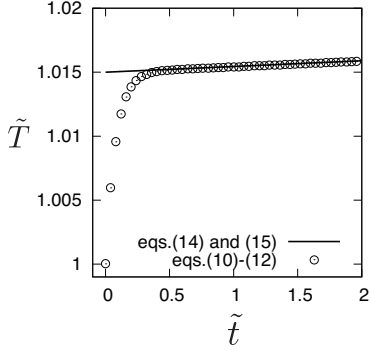}
\caption{Comparison between the dynamics of $\tilde{T}$ obtained as the solution of eqs.~(\ref{eq.theta_nodim})--(\ref{eq.T_nodim}) 
before the adiabatic elimination and $\tilde{T}(\theta,\tilde{\omega})$ in eq.~(\ref{eq.adiabatic_elimination_T}) calculated using the solution of eqs.~(\ref{eq.theta_nodim_2}) and (\ref{eq.ptheta_nodim_2}) 
after the adiabatic elimination for $\Delta \tilde{T}=0.03$. 
The initial conditions at $\tilde{t}=0$ are $(\theta, \tilde{\omega},\tilde{T})=(0,0.03,1)$ and $(\theta, \tilde{\omega})=(0,0.03)$, respectively, and $\tilde{T}(0,0.03)=1.015$.
}\label{T_dynamics}
\end{center}
\end{figure}

\begin{figure}
\begin{center}
\includegraphics[scale=0.6]{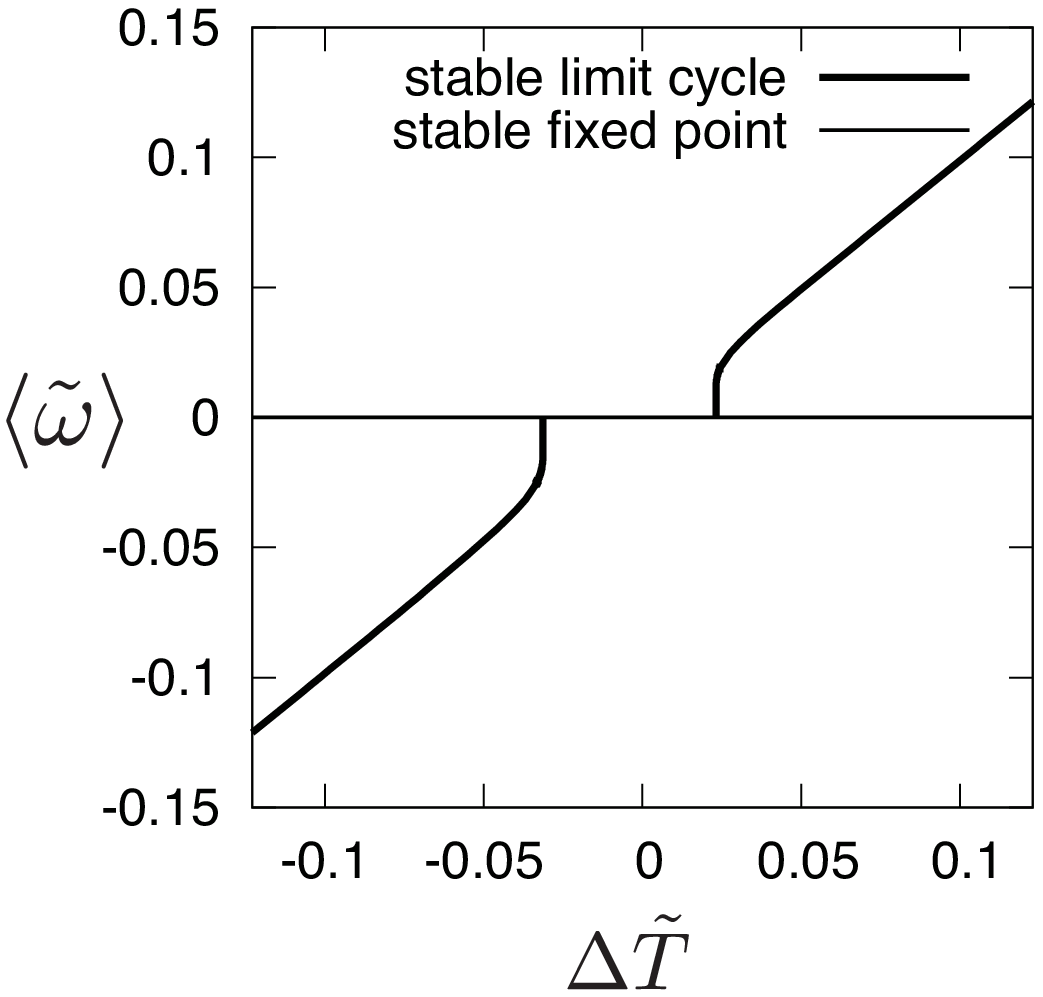}
\end{center}
\caption{$\left<\tilde{\omega}\right>$--$\Delta \tilde{T}$ diagram obtained by solving eqs.~(\ref{eq.theta_nodim_2}) and (\ref{eq.ptheta_nodim_2}). 
}\label{velocity}
\end{figure}

In fig.~\ref{T_dynamics}, we show the dynamics of $\tilde{T}$ obtained as the solution of eqs.~(\ref{eq.theta_nodim})--(\ref{eq.T_nodim})
before the adiabatic elimination for $\Delta \tilde{T}=0.03$ with the initial condition $(\theta, \tilde{\omega},\tilde{T})=(0,0.03,1)$ at $\tilde{t}=0$. 
We also show $\tilde{T}(\theta,\tilde{\omega})$ in eq.~(\ref{eq.adiabatic_elimination_T}) calculated using the solution of eqs.~(\ref{eq.theta_nodim_2}) and (\ref{eq.ptheta_nodim_2}) after the adiabatic elimination with the initial condition $(\theta, \tilde{\omega})=(0,0.03)$ at $\tilde{t}=0$. 
We can see that $\tilde{T}$ quickly relaxes to the adiabatic solution eq.~(\ref{eq.adiabatic_elimination_T}) as expected, 
validating our approximation to use eqs.~(\ref{eq.theta_nodim_2}) and (\ref{eq.ptheta_nodim_2}) instead of eqs.~(\ref{eq.theta_nodim})--(\ref{eq.T_nodim}).

In fig.~\ref{velocity}, we show $\left<\tilde{\omega} \right>$ obtained by 
solving eqs.~(\ref{eq.theta_nodim_2}) and (\ref{eq.ptheta_nodim_2}) 
as a function of the bifurcation parameter $\Delta \tilde{T}$. 
The two attractors---the stable limit cycle and the stable fixed point--- correspond to the rotational state 
with $\left<\tilde{\omega} \right>\ne 0$ and 
the stationary state with $\left<\tilde{\omega} \right>=0$, respectively. 
These states coexist for the sufficiently large $|\Delta \tilde{T}|$, while 
for the too-small $|\Delta \tilde{T}|$, the rotational state does not exist.
The two bifurcation points where the stable limit cycle disappears are
$\Delta \tilde{T}_c \simeq 0.023123$ and $\Delta \tilde{T}'_c\simeq -0.031305$.
The diagram is not symmetric as $\Delta \tilde{T}_c'\ne -\Delta \tilde{T}_c$  
because eqs.~(\ref{eq.theta_nodim_2}) and (\ref{eq.ptheta_nodim_2}) are not invariant under the transformation $\Delta \tilde{T} \to -\Delta \tilde{T}$, $\theta \to -\theta$ and $\tilde{\omega}\to -\tilde{\omega}$.
For $\Delta \tilde{T}_c <\Delta \tilde{T}$, the engine rotates clockwise while for $\Delta \tilde{T}<\Delta \tilde{T}'_c$, 
it rotates counterclockwise, reproducing the actual feature of the engine~\cite{S}.
While the basin structure yielding the stable limit cycle may be complicated, 
the initial angular velocities with the sufficiently large magnitude 
in the expected rotational direction of the engine depending on the sign of $\Delta \tilde{T}$ always lead to the rotational states without reaching the stationary states.

\begin{figure}
\begin{center}
\includegraphics[scale=0.65]{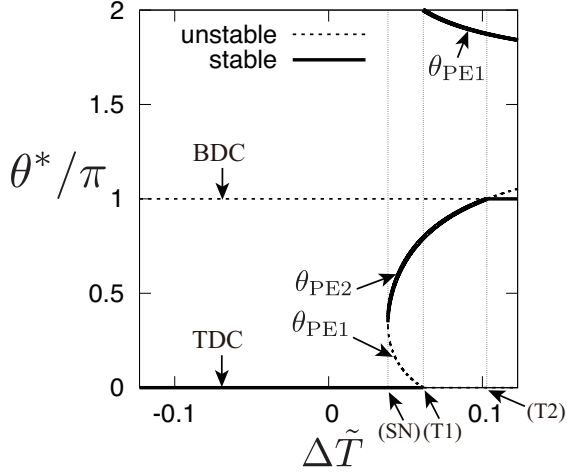}
\caption{$\Delta \tilde{T}$--dependence of $\theta^*=\theta_{{\rm PE}1}, \theta_{{\rm PE}2}$ 
obtained as solutions of the pressure equilibrium 
$\tilde{p}(\theta^*,0)-\tilde{p}_{\rm air}=0$ in eq.~(\ref{eq.ptheta_nodim_2}) shown together with $\theta^*=0({\rm TDC}), \pi({\rm BDC})$.
The solid (dashed) lines and curves
indicate that ${\bm x}^*$ is stable (unstable).
}\label{fixed_point}
\end{center}
\end{figure}

\begin{figure}
\begin{center}
\includegraphics[scale=0.66]{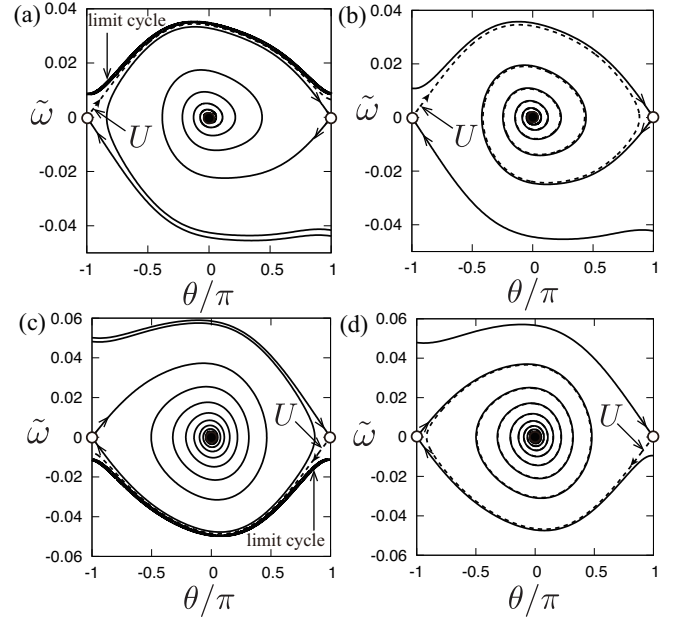}
\caption{The stable spiral point ${\bm x}_0^*$, the saddle point ${\bm x}_\pi^*$, the stable and the unstable manifolds $W^s({\bm x}_\pi^*)$ and $W^u({\bm x}_\pi^*)$ of ${\bm x}_\pi^*$, and the stable limit-cycle orbit
shown on the phase plane for (a) $\Delta \tilde{T}=0.025$, (b) $\Delta \tilde{T}=0.021$, (c) $\Delta \tilde{T}=-0.0345$, and (d) $\Delta \tilde{T}=-0.029$.}\label{homoclinic_bifurcation}
\end{center}
\end{figure}

\section{Fixed-point analysis}
As a preparation to explain the bifurcations in fig.~\ref{velocity},
we investigate fixed points ${\bm x}^*\equiv (\theta^*,\tilde{\omega}^*)$ of eqs.~(\ref{eq.theta_nodim_2}) and (\ref{eq.ptheta_nodim_2}). 
There are four total fixed points at most.

The first two fixed points are the stationary solutions ${\bm x}_0^*\equiv \left(0, 0\right)$ and 
${\bm x}_\pi^*\equiv \left(\pi, 0\right)$ 
obtained from $\sin \theta^*=0$ in eq.~(\ref{eq.ptheta_nodim_2}), 
where $\theta^*$'s are TDC and BDC. 
From the linearized equation 
around ${\bm x}_{0,\pi}^*$,
we obtain the eigenvalue
$\lambda^{\pm}_{0,\pi}=\frac{1}{2}\left(\tau_{0,\pi}\pm \sqrt{\tau_{0,\pi}^2-4\Delta_{0,\pi}}\right)$ in a form using the determinant $\Delta_{0,\pi}$ and the trace $\tau_{0,\pi}$ of the linearized matrix~\cite{Stz}, 
which are given by
\begin{eqnarray}
&&\tau_{0,\pi}=-\tilde{\Gamma}, \ \Delta_0=-\tilde{\sigma} \left(\frac{1}{2}-\tilde{p}_{\rm air}+\frac{\Delta \tilde{T}}{4}\right),\\ 
&&\Delta_\pi=-\tilde{\sigma}\Biggl[\tilde{p}_{\rm air}-\frac{1}{2(1+\tilde{\sigma})}-\frac{\Delta \tilde{T}}{4(1+\tilde{\sigma})}\Biggr].
\end{eqnarray}
We note that ${\bm x}_0^*$ and ${\bm x}_\pi^*$
always exist for any parameter.

The other two fixed points represent the stationary solutions ${\bm x}_{{\rm PE}1}^*\equiv (\theta_{{\rm PE}1},0)$ and
${\bm x}_{{\rm PE}2}^*\equiv (\theta_{{\rm PE}2},0)$
obtained from the pressure equilibrium $\tilde{p}(\theta^*,0)-\tilde{p}_{\rm air}=0$ in eq.~(\ref{eq.ptheta_nodim_2}).
These solutions may not exist depending on the parameters.
Because it is difficult to obtain $\theta_{{\rm PE}1}$ and $\theta_{{\rm PE}2}$ analytically, 
we obtain their $\Delta \tilde{T}$ dependence numerically with the other parameters being fixed in fig.~\ref{fixed_point}. 
The determinant and the trace of the linearized matrix 
around ${\bm x}_{{\rm PE}k}^*$ ($k=1,2$)
are given by
\begin{eqnarray}
&&\Delta_{{\rm PE}k}=-\frac{\tilde{\sigma} \Delta \tilde{T}\sin \theta_{{\rm PE}k} \cos \theta_{{\rm PE}k}-2\tilde{\sigma}^2 \tilde{p}_{\rm air}\sin^2 \theta_{{\rm PE}k}}{2\tilde{V}(\theta_{{\rm PE}k})},\\
&&\tau_{{\rm PE}k}=-\frac{\tilde{\sigma}^2\tilde{p}_{\rm air}\sin^2 \theta_{{\rm PE}k}}{\tilde{G}\tilde{V}(\theta_{{\rm PE}k})}-\tilde{\Gamma},
\end{eqnarray}
respectively.
By using the data of $\theta_{{\rm PE}k}$ obtained in fig.~\ref{fixed_point}, 
we can determine the $\Delta \tilde{T}$ dependence of $\Delta_{{\rm PE}k}$ and $\tau_{{\rm PE}k}$ (data not shown). 
We can then classify the type of 
${\bm x}_{{\rm PE}1}^*$ and ${\bm x}_{{\rm PE}2}^*$ together with that of ${\bm x}_0^*$ and ${\bm x}_\pi^*$
as follows (see also fig.~\ref{fixed_point})~\cite{Stz}.
For $\Delta \tilde{T}<0$, ${\bm x}_{{\rm PE}1}^*$ and ${\bm x}_{{\rm PE}2}^*$ do not exist because 
there does not exist any solution satisfying the pressure equilibrium,
while ${\bm x}_0^*$ and ${\bm x}_\pi^*$ are a stable spiral point and a saddle point, respectively, 
as confirmed from $\Delta_0>0$, $\tau_0<0$ and $\tau_0^2-4\Delta_0<0$ for ${\bm x}_0^*$ and $\Delta_\pi<0$ for ${\bm x}_\pi^*$.
As $\Delta \tilde{T}$ increases, ${\bm x}_{{\rm PE}1}^*$ and ${\bm x}_{{\rm PE}2}^*$ appear via a saddle node bifurcation at $\Delta \tilde{T}\simeq 0.03854$ 
($0<\theta_{{\rm PE}1}=\theta_{{\rm PE}2}<\pi$) ((SN) in fig.~\ref{fixed_point}).
${\bm x}_{{\rm PE}2}^*$, the stable node, immediately changes  
to the stable spiral point at $\Delta \tilde{T}$ satisfying $\tau_{{\rm PE}2}^2-4\Delta_{{\rm PE}2}=0$.
As $\Delta \tilde{T}$ further increases, ${\bm x}_0^*$, 
the stable spiral point, changes to the stable node at $\Delta \tilde{T}=2(2\tilde{p}_{\rm air}-1)-\frac{\tilde{\Gamma}^2}{\tilde{\sigma}}\simeq 0.06156$ obtained from $\tau_0^2-4\Delta_0=0$.
At $\Delta \tilde{T}=2(2\tilde{p}_{\rm air}-1)\simeq 0.06164$ obtained from $\Delta_0=0$, $\theta_{{\rm PE}1}$ agrees with $0$ (TDC, identified with $2\pi$) 
and ${\bm x}_{{\rm PE}1}^*$ and ${\bm x}_0^*$ change to the stable node and the saddle point, 
respectively, switching their stabilities via a transcritical bifurcation ($0<\theta_{{\rm PE}2}< \pi < \theta_{{\rm PE}1}=2\pi$) ((T1) in fig.~\ref{fixed_point}).
As $\Delta \tilde{T}$ increases from this point, ${\bm x}_{{\rm PE}1}^*$ immediately changes from the stable node to the stable spiral point.
As $\Delta \tilde{T}$ further increases, ${\bm x}_{{\rm PE}2}^*$ changes from the stable spiral point to the stable node once again before
achieving the condition
$\Delta \tilde{T}=4\tilde{p}_{\rm air}(1+\tilde{\sigma})-2\simeq 0.1029$ obtained from $\Delta_\pi=0$.
At this $\Delta \tilde{T}$, $\theta_{{\rm PE}2}$ agrees with $\pi$ (BDC) and 
${\bm x}_{{\rm PE}2}^*$ and ${\bm x}_\pi^*$ change to the saddle point and the stable node, 
respectively, switching their stabilities via a transcritical bifurcation ($0< \theta_{{\rm PE}2} \le \pi <\theta_{{\rm PE}1}$) ((T2) in fig.~\ref{fixed_point}). 

As understood from this analysis, at least one stable fixed point always exists, which explains the stationary state with $\left<\tilde{\omega}\right>=0$ in fig.~\ref{velocity}.

\section{Homoclinic bifurcation}
We show that the disappearance of the stable limit cycle 
at $\Delta \tilde{T}_c$ and $\Delta \tilde{T}_c'$  
in fig.~\ref{velocity} occurs via a homoclinic bifurcation as a global bifurcation~\cite{Stz,GH}, where the saddle point ${\bm x}_\pi^*$ plays an essential role in this bifurcation.

In fig.~\ref{homoclinic_bifurcation} (a) and (b),
the stable spiral point ${\bm x}_0^*$ and saddle point ${\bm x}_\pi^*$  
are shown on the phase plane for $\Delta \tilde{T}=0.025$ and $\Delta \tilde{T}=0.021$ before and after the bifurcation at $\Delta \tilde{T}_c\simeq 0.023123$ in fig.~\ref{velocity}.
${\bm x}_{{\rm PE}1}^*$ and ${\bm x}_{{\rm PE}2}^*$ do not exist in these $\Delta \tilde{T}$'s as seen from fig.~\ref{fixed_point}.
The stable and the unstable manifolds $W^s({\bm x}_\pi^*)$ and $W^u({\bm x}_\pi^*)$ of ${\bm x}_\pi^*$ and the stable limit-cycle orbit with $\left<\tilde{\omega}\right>>0$ are also shown on the phase plane.
Here, $W^s({\bm x}_\pi^*)$ ($W^u({\bm x}_\pi^*)$) of ${\bm x}_\pi^*$ is defined as the set of initial values of ${\bm x}(\tilde{t})$
satisfying $\lim_{\tilde{t}\to \infty}{\bm x}(\tilde{t})={\bm x}_\pi^*$ ($\lim_{\tilde{t}\to -\infty}{\bm x}(\tilde{t})={\bm x}_\pi^*$)~\cite{Stz}.
The bifurcation scenario is then described as follows~\cite{Stz}.
In fig.~\ref{homoclinic_bifurcation} (a), we can see that the branch of $W^u({\bm x}_\pi^*)$ labeled $U$ 
(the dashed curve) converges to the stable limit-cycle orbit (the bold curve) as $\tilde{t}\to \infty$.
As $\Delta \tilde{T}$ decreases, these two orbits come close to each other. Finally, at $\Delta \tilde{T}=\Delta \tilde{T}_c$, the limit-cycle orbit touches the saddle point ${\bm x}_\pi^*$,
forming a homoclinic orbit connecting ${\bm x}_\pi^*$ to itself in an infinite period.
At the same time, the limit-cycle orbit disappears.
As $\Delta \tilde{T}$ further decreases, 
$U$ becomes a heteroclinic orbit connecting ${\bm x}_\pi^*$ and ${\bm x}_0^*$.
The same scenario applies for the disappearance of the stable limit-cycle orbit with $\left<\tilde{\omega}\right><0$ at $\Delta \tilde{T}'_c\simeq -0.031305$ in fig.~\ref{velocity}.
Compare the behavior of $U$ 
for $\Delta \tilde{T}=-0.0345$ and $\Delta \tilde{T}=-0.029$ before and after the bifurcation in fig.~\ref{homoclinic_bifurcation} (c) and (d).

This bifurcation is physically understood as follows~\cite{Stz}. As $|\Delta \tilde{T}|$ decreases,  
it becomes increasingly difficult for the crank to sustain the rotational motion by overcoming BDC against the frictional torque 
even with the aid of the moment of inertia.
At the critical value, the crank with vanishingly small speed at BDC takes an infinitely long time to pass and then return to it, which corresponds to the formation of the homoclinic orbit.
The crank can no longer sustain the rotational motion for the smaller $|\Delta \tilde{T}|$.

\section{Discussion}
Our model captures the qualitative features of actual LTD kinematic Stirling heat engines 
and predicts the homoclinic bifurcation.
Accordingly, comparison with experiments must be important to confirm the validity of the model.
The bifurcation points 
$\Delta \tilde{T}_c \simeq 0.023123$ and $|\Delta \tilde{T}'_c| \simeq 0.031305$ in fig.~\ref{velocity} correspond to
$\Delta T_c\simeq 6.775 \ {\rm K}$ and $|\Delta T_c'| \simeq 9.172 \ {\rm K}$ in the original parameters.
These values less than $10 \ {\rm K}$ may be comparable to measured values at which the actual engines including the N-92 type begin rotating~\cite{S}.
The model thus reproduces the actual engine behavior quantitatively to some extent, as well as qualitatively.
However, it seems that experiments that aim to investigate a type of a bifurcation in LTD kinematic engines 
have not yet been conducted.
Because our bifurcation scenario is a mathematically natural consequence of the physically sound model, 
it is important to verify this scenario experimentally.
Besides, experimental evaluations 
of the extent to which this minimal model oversimplifies actual engines will also be important, especially for an appropriate extension of the model.

\section{Conclusion}
We proposed a minimal nonlinear dynamic model of the LTD kinematic Stirling heat engine.
By replacing the role of the displacer piston with the state-dependent thermal conductance
and eliminating the thermodynamic degree of freedom adiabatically,
we derived our dynamical equations that describe the engine as the driven nonlinear pendulum.
Our model reproduced the rotational motion of the engine with the expected direction as the stable limit cycle and the stationary state of the engine as the stable fixed point, respectively.
We clarified that the stable limit cycle disappears via the homoclinic bifurcations, with the temperature difference being the bifurcation parameter.
Some of the topics that we could not discuss in this study, such as an analysis of the performance of the engine, will be reported elsewhere.
Although we have considered the simplest case of the kinematic Stirling heat engine, 
it is of great interest to investigate a bifurcation scenario of a Ringbom engine~\cite{RGK,S2} as the other LTD Stirling heat engine~\cite{KW} based on nonlinear dynamics.
Moreover, we may even invent completely new types of LTD heat engines by using nonlinear dynamics as a powerful design and analytical tool. 
We expect that the present work will contribute to deepening the understanding of the physics of the Stirling heat engine and facilitate the use of Stirling heat engines for the development of sustainable societies.

\acknowledgments
The author is grateful to H. Kori for helpful discussions.
This work was supported by JSPS KAKENHI Grant Number 16K17765.

\end{document}